\documentclass[aps,amssymb,floatfix,showpacs]{revtex4}
\usepackage{bm}
\usepackage{amsmath}
\usepackage{graphicx}
\usepackage{float}
\usepackage{setspace}
\usepackage{subfig}
\begin{document}
\def\ba{{\bf a}}
\def\bk{{\bf k}}
\def\bp{{\bf p}}
\def\bq{{\bf q}}
\def\br{{\bf r}}
\def\bv{{\bf v}}
\def\bx{{\bf x}}
\def\bP{{\bf P}}
\def\bR{{\bf R}}
\def\bK{{\bf K}}
\def\bJ{{\bf J}}
\def\la{\langle}
\def\ra{\rangle}
\def\beq{\begin{equation}}
\def\eeq{\end{equation}}
\def\bea{\begin{eqnarray}}
\def\eea{\end{eqnarray}}
\def\bdm{\begin{displaymath}}
\def\edm{\end{displaymath}}
\def\nn{\nonumber}

 \title{Persistent currents in a bosonic mixture in the ring
geometry}

\author{K. Anoshkin}
\affiliation{Department of Physics, Engineering Physics, and 
Astronomy, Queen's University, Kingston, ON K7L 3N6, Canada}

\author{Z. Wu}
\affiliation{Department of Physics, Engineering Physics, and 
Astronomy, Queen's University, Kingston, ON K7L 3N6, Canada}

\author{E. Zaremba}
\affiliation{Department of Physics, Engineering Physics, and 
Astronomy, Queen's University, Kingston, ON K7L 3N6, Canada}

\date{\today}

\begin{abstract}
In this paper we analyze the possibility of persistent currents
of a two-species bosonic mixture in the one-dimensional 
ring geometry. We extend the arguments used by
Bloch~\cite{Bloch73} to obtain
a criterion for the stability of persistent currents for the
two-species system. If the
mass ratio of the two species is a rational number, persistent
currents can be stable at multiples of a certain total
angular momenta. We show that the Bloch criterion can also be
viewed as a Landau criterion involving the elementary
excitations of the system. Our analysis reveals that persistent
currents at higher angular momenta are more stable for the
two-species system than previously thought.
\end{abstract}


\maketitle
\section{introduction}

The hallmark of superfluidity is the possibility of dissipationless 
flow in situations where the flow of a normal fluid would
degrade as a result of viscosity. The textbook example of this
is the flow of a superfluid through a narrow
capillary~\cite{Wilks70}. According to the Landau 
criterion~\cite{Lifshitz80}, the superfluid component
flows without dissipation provided the
superfluid velocity does not exceed some critical value. In this
situation, the normal component remains locked to the walls of
the capillary whereas the superfluid component, carrying zero
entropy, flows as if the
walls of the capillary behaved as a perfectly smooth conduit.
If the capillary is now bent into a torus, one can imagine
that a flow, once established, could persist indefinitely.

The conditions under which persistent currents can occur for a
bosonic mixture in the ring geometry is the subject of this
paper. The usual analysis~\cite{Lifshitz80} leading to the 
Landau criterion is not
obviously applicable since one cannot
invoke Galilean invariance for this closed system. However, for
a system having a single component, Bloch~\cite{Bloch73} presented 
general arguments based on an analysis of the quantum mechanical 
many-body wave function which provided a criterion for
persistent currents. He considered an
idealized one-dimensional ring of radius $R$ in which the particles
interact via an arbitrary pair-wise interaction. Since the total
angular momentum  commutes with the Hamiltonian of the system,
the stationary states have energies $E_\alpha(L)$ which are
functions of the angular momentum quantum number $L$; all other
quantum numbers are subsumed in the index $\alpha$. Bloch
showed that these energy eigenvalues take the form
\beq
E_\alpha(L)=\frac{L^2}{2M_{\rm T}R^2}+e_\alpha(L)
\label{spec_E}
\eeq
where $M_{\rm T}=NM$ is the total mass of the system containing $N$
particles of mass $M$.
The first term on the right hand side of Eq.~(\ref{spec_E}) is
interpreted as the kinetic energy of a rigid ring rotating with
angular velocity $\Omega = L/M_TR^2$. The second term,
$e_\alpha(L)$, corresponds 
to internal excitations of the system; it has the periodicity
property
\beq
e_\alpha(L+N\hbar)=e_\alpha(L).
\label{e_period}
\eeq
This implies that the system can
find itself in the same internal state for angular momenta that
differ from each other by multiples of $N\hbar$. In addition,
$e_\alpha(L)$ has the inversion property
\beq
e_\alpha(-L)=e_\alpha(L),
\label{e_inversion}
\eeq
which reflects the fact that the energy does not depend on the
sense of the angular momentum.

The state with the lowest energy for a given $L$ will be given
the label $\alpha = 0$. In the noninteracting limit,
$e_0(L)$ has a local minimum at $L=0$~\cite{Bloch73}; one expects 
this property to persist with repulsive interactions. The
periodicity of this function then implies that $E_0(L)$ can
exhibit local minima at certain multiples of $N\hbar$. If and
when such minima occur, Bloch argued that the system is capable
of sustaining persistent currents.
Conversely, if $E_0(L)$ is not at a local minimum,
nonidealities will induce transitions which change the angular
momentum  and hence the flow of the superfluid around the ring.


In Sec.~\ref{Bloch}, we extend Bloch's analysis to a two-species
gas containing $N_A$ particles of type $A$ and $N_B$ particles
of type $B$. Here the term ``species'' can
refer either to different kinds of atoms or to atoms
distinguished by their hyperfine states. 
When the masses of the two species are
different, we find that the energy can still be written
in the form of Eq.~(\ref{spec_E}) but in general, $e_0(L)$
is no longer
a periodic function of $L$. However, if the masses are equal,
$e_0(L)$ is found to have the same periodicity as for the
single-species case with $N=N_A+N_B$. In the case that the mass 
ratio $M_A/M_B$
is a rational number, $e_0(L)$ remains a periodic function
of $L$ but with a periodicity that differs from $N\hbar$. For
these special cases, Bloch's arguments for the possibility of
persistent currents goes through as for the single-species case. 
For arbitrary mass ratios, $E_0(L)$ may still exhibit a local
minimum at some finite value of $L$ but there is no general 
argument which can be used to determine where such a local 
minimum might occur.

We go on to show that Bloch's criterion for persistent currents
can be phrased in terms of the more familiar Landau criterion.
For $M_A = M_B$, $e_0(L)$ is periodic and
a Landau criterion can be formulated at the discrete set of angular
momenta $L = L_n = n N\hbar$, with $n$ an integer, where
the system can be taken to be in its internal 
ground state. The Landau criterion then imposes a constraint on
the spectrum of the elementary excitations with angular momentum
$m\hbar$ and energy $\varepsilon(m)$. In Sec.~\ref{Bogoliubov},
these excitation energies are determined for the two-species
system in the Bogoliubov approximation. In general there are two
Bogoliubov modes which are usually phonon-like at long
wavelengths. For the case $M_A = M_B$, the Landau criterion then 
suggests
that persistent currents may be stable for certain values of $n$.
However, if the interaction parameters satisfy
a certain relation (given in Sec.~\ref{Bogoliubov}),
one of the Bogoliubov modes has a particle-like
dispersion and the Landau criterion leads to the conclusion that
persistent currents are unstable for all $n$.

The above conclusion was arrived at earlier by Smyrnakis {\it et
al.}~\cite{Smyrnakis09} based on an analysis of
the mean-field Gross-Pitaevskii (GP)
energy functional for the two-species system. With the
assumption that all interaction parameters are equal, these
authors determine $E_0(L)$ by minimizing the GP energy functional
subject to the constraint that the average angular momentum of
the system is $L$. Although persistent currents are destabilized
at $L=L_n$, the authors find that $E_0(L)$ can exhibit local minima
at non-integral values of $l = L/N\hbar$. In particular, they
show that persistent currents are stable at $l = x_A =
N_A/(N_A+N_B)$, provided the interactions are sufficiently
strong. Furthermore, their analysis leads to the conclusion that
persistent currents are unstable for $l>1$ even when the
concentration of the minority component is arbitrarily small.
This latter conclusion seems at odds with what one might expect
in the pure single-species limit ($x_B \to 0$).

In Sec.~\ref{GP-Theory}, we present the analysis of the GP
energy functional in somewhat more detail than was provided by
Smyrnakis {\it et al.}~\cite{Smyrnakis09} This analysis essentially confirms all of
their analytical results, however, we find that the information
regarding the behaviour of
$e_0(L)$ in the vicinity of $L = N_A\hbar$ is not sufficient to
establish whether or not persistent currents are actually
stable. In fact, a more global analysis of $e_0(L)$ shows that
persistent currents can exist when $l>1$. Our work also
clarifies how the single-species results are recovered in the
$x_B \to 0$ limit.

\section{Bloch's criterion for persistent currents in a 
two-species gas}
\label{Bloch}
In this section we extend
Bloch's analysis to a two-species system consisting of $N_A$
particles of type $A$ and $N_B$ particles of type 
$B$. The masses of the particles are $M_A$ and $M_B$. In
addition, we assume an idealized one-dimensional ring geometry.
The Hamiltonian $H$ for this system is taken to be
\begin{equation}
\label{Two-comp hamiltonian}
{H} = \sum\limits_{i=1}^{N_A}\frac{\hat{l}_i^2}{2M_AR^2} + 
\sum\limits_{i=N_A+1}^{N_A+N_B}\frac{\hat{l}_i^2}{2M_BR^2}
+\sum\limits_{i<j}v_{ij}(\theta_i - \theta_j),
\end{equation}
where the angular momentum operator of the $i$-th particle 
about the centre of the ring is
\begin{equation}
\label{ang_momentum_ring}
\hat{l}_i = \frac{\hbar}{i}\frac{\partial}{\partial \theta_i}.
\end{equation}
The index $i$ denotes an $A$-type particle for $1\le i\le N_A$
and a $B$-type particle for $N_A+1 \le i \le N_A+N_B \equiv N$.
The subscripts on the interaction potential $v_{ij}$ allow for
the interactions between the particles to be species-dependent.
For the pair-wise interactions assumed, the total angular
momentum
\begin{equation}
\label{total_ang_momentum}
\hat{L} = \sum\limits_{i=1}^{N}\hat{l}_i = \sum\limits_{i=1}^N \frac{
\hbar}{i}\frac{\partial}{\partial \theta_i}
\end{equation}
commutes with the Hamiltonian. The stationary states 
$\Psi(\theta_1,...,\theta_N)$ of the Hamiltonian can thus be chosen 
to be simultaneous eigenstates of the total angular momentum. 

A suitable basis of states can be constructed from the following
product states for noninteracting particles:
\begin{equation}
\label{wave_function_Phi}
\Phi(\theta_1, ..., \theta_N) =  
\phi_{m_1}(\theta_1)\phi_{m_2}(\theta_2)\cdots
\phi_{m_N}(\theta_N).
\end{equation}
Here $m_i$ is an integer and
\beq
\phi_m(\theta) = \frac{e^{im\theta}}{\sqrt{2\pi}}.
\label{s_p_basis}
\eeq
The wave function in
Eq.~(\ref{wave_function_Phi}) is an eigenfunction of $\hat L$ with 
eigenvalue $L = \hbar \sum_i m_i$.  It can be written in
different ways. One possibility is
\begin{equation}
\label{wave_function_Phi_II}
\Phi(\theta_1, ..., \theta_N) = (2\pi)^{-N/2}
\exp (iNl\Theta )\exp \bigg [\frac{i}{N}\sum_{ij}
m_i(\theta_i-\theta_j)\bigg]
\end{equation}
where
\begin{equation}
l = \frac{1}{N} \sum_{i=1}^N m_i
\end{equation}
is the angular momentum per particle in units of $\hbar$ and
\begin{equation}
\Theta = \frac{1}{N}\sum_{i=1}^N \theta_i
\end{equation}
is the mean angular coordinate. The above wave function is
identical in form to that of a single species system.
By construction, the first exponential in Eq. 
(\ref{wave_function_Phi_II}) is an eigenfunction of
$\hat L$ with eigenvalue $L = Nl\hbar$. The second exponential
is a function of
the coordinate differences $\theta_i-\theta_j$ and as such
is a zero total angular momentum wave function.

Properly symmetrized functions are obtained from Eq. 
(\ref{wave_function_Phi_II}) with the
application of the symmetrization operator
\begin{equation}
\label{symmetrization_two-species}
\hat{S} = \hat{S}_A\hat{S}_B
\end{equation}
where
\begin{eqnarray}
\label{Symm_A}
&&\hat{S}_A = \frac{1}{N_A!}\sum\limits_{P_A}\hat{P}_A \\
\label{Symm_B}
&&\hat{S}_B = \frac{1}{N_B!}\sum\limits_{P_B}\hat{P}_B. 
\end{eqnarray}
The operator $\hat{P}_A$ permutes the coordinates of the $A$ particles, 
whereas $\hat{P}_B$ does the same for $B$ particles. Applying the 
symmetrization operator to the wave function $\Phi(\theta_1, ..., 
\theta_{N})$ yields
\begin{equation}
\label{Phi_two-species}
\Phi_{\{m_i\}}( \theta_1, ..., \theta_N ) = \exp\left( iNl\Theta 
\right) \tilde \chi_{\{m_i\}}(\theta_1, ..., \theta_N),
\end{equation}
where $ \tilde \chi_{\{m_i\}}$ is a normalized function of the 
coordinate differences $\theta_i - \theta_j$. The 
functions in Eq.~(\ref{Phi_two-species}) provide a basis of properly
symmetrized $N$-particle states, with $N= N_A+N_B$.

The stationary state solutions of $\hat H \Psi = E\Psi$ with
angular momentum $L$ will be denoted
$\Psi_{L\alpha}(\theta_1,...,\theta_N)$, where $\alpha$ indicates 
the rest of the
quantum numbers.
These states can be expanded in terms of the basis functions 
Eq.~(\ref{Phi_two-species}) as
\begin{eqnarray}
\label{Psi_L}
\nonumber
&&\Psi_{L\alpha}(\theta_1, ..., \theta_N) = {\sum_{\{m_i\}}}'
C_{L\alpha}(\{m_i\})\Phi_{\{m_i\}}(\theta_1, ..., \theta_N)\\
&&\hspace{2.15cm}\equiv \exp\left[ iNl\Theta \right] 
\tilde \chi_{L\alpha}(\theta_1, 
..., \theta_N),
\end{eqnarray}
where the prime on the summation implies the restriction
$\sum_{i=1}^{N}m_i=Nl$.
It is clear from the way $\tilde \chi_{L\alpha}(\theta_1,...,
\theta_N)$ is defined that it is a function of 
the relative angular coordinates $\theta_i-\theta_j$. 
Substituting Eq. (\ref{Psi_L}) into the Schr\"odinger equation for 
$\Psi_{L\alpha}$, we
find that $\tilde \chi_{L\alpha}$ satisfies the equation
\begin{equation}
\label{chi_L_SE}
H_L\tilde \chi_{L\alpha} = \tilde e_\alpha(L)\tilde \chi_{L\alpha},
\end{equation}
where 
\begin{equation}
H_L = H +
\frac{L}{NR^2}\left(\frac{\hat{L}_A}{M_A}+ 
\frac{\hat{L}_B}{M_B}\right)
\end{equation}
and
\begin{equation}
\label{epsilon of L}
\tilde e_\alpha(L) = E_\alpha(L) - \left( \frac{\hbar^2 l^2 N_A}{2M_AR^2} + 
\frac{\hbar^2 l^2 N_B}{2M_BR^2} \right) = E_\alpha(L) - 
\frac{L^2}{2N^2R^2}\left( 
\frac{N_A}{M_A} + \frac{N_B}{M_B} \right).
\end{equation}
Since $\hat L \tilde \chi_{L\alpha} = 0$, $H_L$ in 
Eq.~(\ref{chi_L_SE}) can be expressed equivalently as
\begin{equation}
\label{H_L}
H_L = H +
\frac{L}{NR^2}\left(\frac{1}{M_A}- 
\frac{1}{M_B}\right )\hat L_A.
\end{equation}
We observe that this Hamiltonian is in general $L$-dependent
which has important consequences for $\tilde e_\alpha(L)$.

Eq.~(\ref{chi_L_SE}) must be solved with appropriate boundary
conditions. Since the
wave function $\Psi_{L\alpha}$ is required to be single-valued with 
respect to each of the angular variables, it satisfies
\begin{equation}
\label{Periodicity_Psi}
\Psi_{L\alpha}(...,\theta_i+2\pi,...) = \Psi_{L\alpha}(...,\theta_i,...).
 \end{equation}
Eq.~(\ref{Periodicity_Psi}) then implies
\begin{equation}
\label{periodicity_chi}
\tilde \chi_{L\alpha}(..., \theta_i+2\pi, ...) = \exp\left[ 
-i2\pi l \right]\tilde \chi_{L\alpha}( ..., \theta_i, ... )
\end{equation}
for $i = 1, ..., N$. From this we see that the boundary
conditions are periodic as a function of $L$ with period
$N\hbar$. With the basis functions written in the form given in
Eq.~(\ref{wave_function_Phi_II}), they are the {\it same} 
boundary conditions that apply in 
the single-species case. In the $N_B = 0$ limit, $\tilde
e_\alpha(L) = E_\alpha(L) - L^2/2M_T R^2 \equiv e_\alpha(L)$
with $M_T = NM_A$. In addition, the
Hamiltonian $H_L$ reduces to $H$ since $\hat L_A \to
\hat L$ and $\hat L \tilde \chi_{L\alpha} = 0$. As a result,
$\tilde \chi_{L'\alpha}$ with
$L' = L+N\hbar$ satisfies the same Schr\"odinger equation and
boundary conditions as $\tilde \chi_{L\alpha}$. This implies that the 
eigenvalue spectrum for these two functions is identical. As
concluded by Bloch~\cite{Bloch73}, the eigenvalues $e_\alpha(L)$ for the
single-component system are then periodic functions of $L$ with
period $N\hbar$. In particular, the ground state energy is given
by
\begin{equation}
E_0(L) = \frac{L^2}{2NM_AR^2} + e_0(L).
\label{E_0}
\end{equation}
The same considerations apply to the two-species situation for
the special case $M_A = M_B$ since $H_L$ in Eq.~(\ref{H_L}) also 
reduces
to $H$ in this limit and Eq.~(\ref{E_0}) is still valid. In both of 
the above situations, 
the periodicity of the eigenvalues $e_\alpha(L)$ means physically 
that the ``internal" excitations can be the same for distinct
macroscopic flows whose angular momenta differ by some
multiple of $N\hbar$. However, $\tilde e_0(L)$ is no longer 
periodic when $M_A \ne M_B$, since the Hamiltonian $H_L$ in
Eq.~(\ref{chi_L_SE}) depends explicitly on $L$.

An alternative analysis is provided by 
writing the wave function in (\ref{Psi_L}) as
\begin{equation}
\label{Psi_L_cm}
 \Psi_{L\alpha}(\theta_1, ..., \theta_N)=\exp (iNl\Theta_{\rm cm})
\chi_{L\alpha}(\theta_1, ..., \theta_N),
\end{equation}
where $\Theta_{\rm cm}$ is the `centre-of-mass' angular coordinate
defined as
\begin{equation}
\Theta_{\rm cm} =\frac{1}{M_T}\sum_{i=i}^N M_i\theta_i
\end{equation}
and 
\begin{equation}
\chi_{L\alpha}(\theta_1, ..., \theta_N)=\exp\left [ 
-i\frac{l}{M_T}\sum_{ij}^N M_i(\theta_i-\theta_j)\right 
]\tilde \chi_{L\alpha}(\theta_1, ..., \theta_N).
\end{equation}
Here and in the following, $M_i$ is equal to $M_A$ for $i\leq N_A$ 
and $M_B$ for $i> N_A$; $M_T = N_A M_A + N_B
M_B$ is the total mass. We observe that the exponential in
Eq.~(\ref{Psi_L_cm}) is still an eigenfunction of $\hat L$ 
with eigenvalue $L$ and
that $\chi_{L\alpha}$ is a function of coordinate
differences and therefore a zero-angular momentum function.
Eq.~(\ref{Psi_L_cm})
amounts to a separation of the centre-of-mass motion from 
the internal degrees of freedom.
Indeed, substitution of Eq.~(\ref{Psi_L_cm}) into the Schr\"odinger 
equation for $\Psi_{L\alpha}$ 
yields


\begin{equation}
\label{tilde_chi_L_SE}
H \chi_{L\alpha} = e_\alpha(L)\chi_{L\alpha},
\end{equation}
where 
\begin{equation}
\label{tilde_eps}
e_\alpha(L)=E_\alpha(L)-\frac{L^2}{2M_TR^2}.
\end{equation}
Eqs.~(\ref{tilde_chi_L_SE}) and (\ref{tilde_eps}) suggest that 
$\chi_{L\alpha}$ and $e_{\alpha}(L)$ can be viewed,
respectively, as the 
``internal'' wave function and ``internal" excitation energy.
The boundary conditions imposed on 
$\chi_{L\alpha}(\theta_1,...,\theta_N)$ can be derived from Eq. 
(\ref{Psi_L_cm}) and are given by
\beq
\label{tilde_chi_bc}
\chi_{L\alpha}(\cdots,\theta_i+2\pi,\cdots)=\exp\left 
(-i2\pi\frac{\nu M_i}{M_{T}}\right 
)\chi_{L\alpha}(\cdots,\theta_i,\cdots),
\eeq
where $\nu=Nl$.
When $M_A=M_B=M$, these boundary conditions revert to those of 
the single-species case where
\beq
\chi_{L\alpha}(\cdots,\theta_i+2\pi,\cdots)=\exp\left (-i2\pi l\right 
)\chi_{L\alpha}(\cdots,\theta_i,\cdots).
\eeq
This, together with Eq.~(\ref{tilde_chi_L_SE}) implies that 
$\chi_{L+N\hbar,\alpha}=\chi_{L\alpha}$ and $
e_\alpha(L+N\hbar)= e_\alpha(L)$. In fact, in this case
$e_\alpha(L)$ and $\chi_{L\alpha}$ coincide with 
$\tilde e_n(L)$ and $\tilde \chi_{L\alpha}$, respectively.

When $M_A\neq M_B$, $e_\alpha(L)$ 
is not in general a periodic function of $L$. However, it can be
if the boundary conditions in Eq.~(\ref{tilde_chi_bc}) remain
unaltered when $\nu$ is augmented by some number $\tilde{N}$ 
(i.e., $L\rightarrow L+\tilde{N}\hbar$) such that 
\beq
\frac{\tilde{N}M_A}{M_{\rm T}}=p,
\eeq
and
\beq
\frac{\tilde{N}M_B}{M_{\rm T}}=q,
\eeq
where $p$ and $q$ are both integers. This implies that $M_A/M_B$
must be equal to the rational number $p/q$. The lowest possible
value of $\tilde N$ is obtained when $p$ and $q$ have no common
divisor and is then given by 
\beq
\tilde {N}=pN_A+qN_B.
\eeq 
With this choice of $\tilde N$, $e_\alpha(L)$ is a periodic
function of $L$ with periodicity $\tilde N \hbar$. In this 
situation, it is possible to impart a definite angular momentum to 
the two-species system without altering its ``internal" state.
For two different atomic species, the mass ratio
$M_A/M_B$ is never strictly a rational number and thus
$e_\alpha(L)$ cannot be strictly periodic. However, if
\beq
M_A/M_B \simeq p/q + \delta
\eeq
where $|\delta| << p/q$, one would expect $e_\alpha(L)$,
by continuity, to be quasi-periodic with a periodicity
close to $(N_Ap+N_Bq)\hbar$. For example, a
mixture of $^{85}{\rm Rb}$ ($A$) and $^{39}{\rm K}$ 
($B$) has a mass ratio
\beq
M_A/M_B  \simeq 2 + 0.07,
\eeq
in which case
the quasi-periodicity of $e_\alpha(L)$ would be $(2N_A+N_B)\hbar$.


In the rest of this section we discuss the close connection between 
Bloch's argument on persistent currents and Landau's criterion for
superfluidity. Our analysis mainly concerns the single-species and 
equal-mass two-species systems, where there is strict periodicity for 
$e_\alpha(L)$. However, it also applies to the two-species system with 
unequal masses, insofar as it is a good approximation to regard 
$e_0(L)$ as quasi-periodic. According to Bloch, persistent
currents can occur at the angular momenta $L_n = n N\hbar$, 
for integral $n$, if $E_0(L)$ has a local minimum at $L= L_n$.
We thus examine the behaviour of $E_0(L)$ in the neighbourhood 
of $L_n$. From Eqs.~(\ref{spec_E}) and (\ref{e_period}) one 
has
\bea
E_0(L_n+\Delta L)&=&\frac{(L_n+\Delta
L)^2}{2M_{T}R^2}+e_0(L_n+\Delta 
L) \nn \\
&=&\frac{L_n^2}{2M_{ T}R^2}+\Omega_n\Delta L+E_0(\Delta L),
\label{E_0(L)}
\eea
where $\Omega_n\equiv L_n/(M_{\rm T}R^2)$ is the angular velocity 
of the centre of mass of the system at $L_n$. This expression
for the energy is analogous to the expression obtained via a
Galilean transformation for a homogeneous system in which an
excitation is produced in the rest frame of the
superfluid~\cite{Lifshitz80}. To
make this correspondence evident, we define the velocity $v_n
\equiv L_n/M_T R$ and write the energy in Eq.~(\ref{E_0(L)}) as
\beq
E_0(L_n+\Delta L)=
\frac{1}{2}M_T v_n^2 + \left (\frac{\Delta L}{R} \right )
v_n+ E_0(\Delta L).
\eeq
The first term on the right hand side is identified as
the kinetic energy of the superfluid moving with velocity
$v_n$. Likewise, the last term is identified as the energy 
of a stationary superfluid containing an excitation with
``momentum" $\Delta L/R$. It should be noted, however, that the
analogy is not complete since for a homogeneous system the
superfluid velocity $v_n$ can take arbitrary values whereas 
for the ring geometry the angular momentum is restricted to the 
discrete values $L_n$. 

With this correspondence in mind, we take
$E_0(\Delta L)$ to be the energy of the system with a single
quasi-particle excitation with angular momentum $\Delta L = 
\hbar m$ and energy $\varepsilon(m)$, i.e.,
\beq
E_0(\Delta L)=E_0(0)+\varepsilon(m).
\eeq
We thus have 
\beq
E_0(L_n+\Delta L)=E_0(L_n)+
\varepsilon(m)+ \Omega_n  \hbar m. 
\eeq
The stability of the state with energy $E_0(L_n)$ is then
assured if the excitations lead to an increase in energy.
In other words, the system will sustain 
persistent currents at $L_n$ for an arbitrary excitation
of the system if
\beq
\varepsilon(m)+\hbar \Omega_n m>0
\label{lc1}
\eeq
for all $m$. Since $\varepsilon(-m) = \varepsilon(m)$, the left
hand side has a minimum for negative values of $m$ and we thus
require
\beq
\Omega_n < \left (\frac{\varepsilon(m)}{\hbar |m|}\right )_{\rm{min}}.
\label{lc2}
\eeq
We have thus shown that Bloch's argument for persistent currents 
in the one-dimensional ring geometry
naturally leads to the more familiar Landau criterion for 
superfluidity.
If $\varepsilon(m)$ has a positive curvature as a function of
$m$, which precludes a roton-like minimum,
the inequality in Eq.~(\ref{lc2}) can be replaced by
\beq
\Omega_n<\frac{\varepsilon(m=1)}{\hbar}.
\label{lc3}
\eeq
It is clear from this expression that the inequality must
eventually fail
when $n$ exceeds some critical value $n_{\rm cr}$.

\section{Bogoliubov excitations, dynamic stability and Persistent 
currents at integer values of angular momentum per particle}
\label{Bogoliubov}
The Landau criterion derived in the previous section focuses
attention on the elementary excitations of the system. 
In this section, we obtain these excitations for a 
two-species gas in a one-dimensional ring geometry in the
Bogoliubov approximation. We then apply 
Eq.~(\ref{lc3}) to discuss persistent currents at integer values of 
angular momentum per particle for an equal-mass two-species system.

In the following, we assume that the particles interact via
contact interactions with strengths $U_{ss'}$, where 
$s,s'=A,B$ specify the species. Using the single-particle basis
in Eq.~(\ref{s_p_basis}), the Hamiltonian in Eq.~(\ref{Two-comp 
hamiltonian}) can be written in the second-quantized form
\bea
\hat H &=&\sum_s\sum_m \epsilon_s \hat 
a^\dag_{s,m}\hat 
a_{s,m}+\sum_{s,s'}\sum_{m,m',n}\frac{U_{ss'}}{4\pi}\hat 
a^\dag_{s,m}\hat a^\dag_{s',n-m}\hat a_{s',m'}\hat a_{s,n-m'},
\eea
where $m$ is the angular momentum quantum number and $\epsilon_s
=\hbar^2m^2/2M_sR^2$. Assuming both species to be Bose-condensed in
the $m=0$ state, the
corresponding Bogoliubov Hamiltonian can be written as
\bea
\label{H_Bog_0}
\hat H_{\rm
Bog}&=&\frac{1}{2}\sum_{s,s'}\sqrt{N_sN_{s'}}
g_{ss'}
+\sum_{s}\sum_{m\neq 0}\left [ (\epsilon_s+g_{ss})\hat 
a^\dag_{s,m}\hat a_{s,m} +\frac{1}{2}g_{ss}\hat a^\dag_{s,m}\hat 
a^\dag_{s,-m}+\frac{1}{2}g_{ss}\hat a_{s,m}\hat a_{s,-m}\right ] 
\nn \\
&&\hskip 1.45truein +\sum_{s\neq s'}\sum_{m\neq 0}g_{ss'}\left [ \hat 
a^\dag_{s,m}\hat a_{s',m}+\frac{1}{2}\hat a^\dag_{s,m}\hat 
a^\dag_{s',-m}+\frac{1}{2}\hat a_{s,m}\hat a_{s',-m}\right ],
\eea
where 
$g_{ss'}= U_{ss'} \sqrt{N_s N_{s'}}/2\pi$. 

The diagonalization of a Hamiltonian similar to Eq.~(\ref{H_Bog_0}) 
for a three-dimensional system was carried out in~\cite{Tommasini03}. 
Here we present a different method 
of determining the Bogoliubov quasiparticle operators.
This is done in three 
steps. First, we perform a Bogoliubov transformation for each of
the species treated individually. The transformation is defined
by
\bea
\label{Bog_tran_0}
&&\hat a_{s,m}=u^{(0)}_{s,m}\hat \beta_{s,m}-v^{(0)}_{s,m}\hat 
\beta^\dag_{s,-m}  \nn \\
&&\hat a_{s,-m}=u^{(0)}_{s,m}\hat 
\beta_{s,-m}-v^{(0)}_{s,m}\hat \beta^\dag_{s,m},
\eea
with
\beq
\label{bare_Bog_amp}
\big ({u^{(0)}_{s,m}} \big )^2= 
\frac{1}{2} \left (\frac{\epsilon_s+g_{ss}}{E_s}+1  \right ) = 
\frac{(E_s+\epsilon_s)^2}{4E_s \epsilon_s},\qquad 
\big ( {v^{(0)}_{s,m}}\big )^2=
\frac{1}{2} \left (\frac{\epsilon_s+g_{ss}}{E_s}-1  \right ) = 
\frac{(E_s-\epsilon_s)^2}{4E_s \epsilon_s},
\eeq
where
\beq
E_s=\sqrt{\epsilon_s^2+2\epsilon_s g_{ss}}.
\label{Bog_freq}
\eeq
$E_s$ is the Bogoliubov excitation energy for independent
components.
Substituting Eq.~(\ref{Bog_tran_0}) into Eq.~(\ref{H_Bog_0}) and 
dropping all constant terms, we obtain
\bea
\label{H_Bog_1}
\hat H_{\rm Bog}
=\sum_{s}\sum_{m\neq 0}E_s\hat\beta^\dag_{s,m}\hat 
\beta_{s,m} +\sum_{s\neq s'}\sum_{m\neq 0}
\tilde g
\left [ \hat 
\beta^\dag_{s,m}\hat \beta_{s',m} +\frac{1}{2}\hat \beta^\dag_{s,m}\hat 
\beta^\dag_{s',-m}+\frac{1}{2}\hat \beta_{s,m}\hat \beta_{s',-m}\right 
],
\eea
where $\tilde g \equiv
\sqrt{\epsilon_A\epsilon_B/E_AE_B}g_{AB}$. The second term in
this Hamiltonian describes the coupling between the Bogoliubov 
excitations defined for each of the species. It is convenient
to write the Hamiltonian (again to within a constant) in the matrix form
\beq
\label{H_Bog_1_matr}
\hat H_{\rm Bog}=\sum_{m>
0}\hat{\bf{\Phi}}^\dag_m\mathcal{M}\hat{\bf{\Phi}}_m,
\eeq
where 
\bea
&&\hat {\bf{\Phi}}_m\equiv (\hat \beta_{A,m}\quad \hat 
\beta^\dag_{A,-m}\quad\hat \beta_{B,m}\quad \hat \beta^\dag_{B,-m} 
)^{\rm T} \nn \\
&&\hat {\bf{\Phi}}^\dag_m\equiv (\hat \beta^\dag_{A,m}\quad \hat 
\beta_{A,-m}\quad\hat \beta^\dag_{B,m}\quad \hat \beta_{B,-m} )
\eea and 
\beq
\mathcal{M} = \left( \begin{array}{cccc}
E_A & 0 & \tilde{g} & \tilde{g} \\
0 & E_A & \tilde{g} & \tilde{g}  \\
\tilde{g} & \tilde{g} & E_B & 0 \\
\tilde{g} & \tilde{g} & 0 & E_B
\end{array} \right).
\eeq

To complete the diagonalization process we introduce the
following transformations
\bea
&&\hat\beta_{s,m}=\tilde{u}^{(+)}_{s,m}\hat 
\beta_{+,m}-\tilde{v}^{(+)}_{s,m}\hat\beta^\dag_{+,-m}+\tilde{u}^{(-)}_{
s,m}\hat \beta_{-,m}-\tilde{v}^{(-)}_{s,m}\hat \beta^\dag_{-,-m} \nn \\
&&\hat\beta_{s,-m}=\tilde{u}^{(+)}_{s,m}\hat 
\beta_{+,-m}-\tilde{v}^{(+)}_{s,m}\hat\beta^\dag_{+,m}+\tilde{u}^{(-)}_{
s,m}\hat \beta_{-,-m}-\tilde{v}^{(-)}_{s,m}\hat \beta^\dag_{-,m},
\eea
where the amplitudes are chosen to be real. 
The Hamiltonian is reduced to the diagonalized form
\beq
\label{H_Bog_2}
\hat H_{\rm Bog}=\sum_{m\neq 0}E_+\hat\beta^\dag_{+,m}\hat 
\beta_{+,m}
+\sum_{m\neq 0}E_-\hat\beta^\dag_{-,m}\hat \beta_{-,m},
\eeq
if the amplitudes satisfy the matrix equation
\beq
\label{Bogo_equ}
\sigma_z\mathcal{M}\tilde{{\bf w}}_{\pm}=\omega_\pm \tilde{{\bf 
w}}_{\pm}
\eeq
with the normalization condition
\beq
\label{Normalization}
\tilde{{\bf w}}^{\rm T}_{\pm}\sigma_z\tilde{{\bf w}}_{\pm}=1.
\eeq
Here, $\tilde{{\bf w}}_{\pm}\equiv 
(\tilde{u}^{(\pm)}_{A,m}\quad -\tilde{v}^{(\pm)}_{A,m}\quad 
\tilde{u}^{(\pm)}_{B,m}\quad -\tilde{v}^{(\pm)}_{B,m})^{\rm T}$
and the matrix $\sigma_z$ is defined as
\beq
\sigma_z=\left ( \begin {array} {cccc}
1&0&0&0 \\
0&-1&0&0 \\
0&0&1&0 \\
0&0&0&-1
                 \end {array}
\right ).
\eeq
It should be noted that Eqs.~(\ref{Bogo_equ}) and 
(\ref{Normalization}) guarantee that the Bose commutation
relations of the new operators $\hat \beta_{+,m}$  and $\hat
\beta_{-,m}$ are preserved.

The Bogoliubov excitation energies $E_\pm$ are
determined by the characteristic equation
\beq
\label{Char_Equ1}
{\rm det}(\sigma_z\mathcal{M}-E \mathcal{I})=0,
\eeq	
which yields
\begin{equation}
\label{Char_Equ2}
\left(E^2-E_A^2 \right) \left(E^2-E_B^2 \right) - 
4\epsilon_A\epsilon_B g_{AB}^2= 0.
\end{equation} 
This quadratic equation in $E^2$ has the two
roots~\cite{Ao00,Pethick08}
\begin{equation}
\label{Bog_spec}
E_{\pm}^2 = \frac{1}{2}\left( E_A^2 + E_B^2 \right) 
\pm \frac{1}{2}\sqrt{\left(E_A^2 + E_B^2 \right)^2
+4\left(4\epsilon_A\epsilon_B g_{AB}^2 
-E_A^2E_B^2\right)}.
\end{equation}
The dispersion of these modes is `phonon-like' for small $m$
($E_\pm \propto |m|$) and `particle-like' for large $m$ ($E_\pm
\propto m^2$). The upper branch has the higher sound speed
and evolves continuously into $\hbar^2m^2/2M_<R^2$ where $M_<$
signifies the smaller of the two masses.

The Bogoliubov excitations of the two-component system 
are dynamically stable provided
$E^2 >0$. Since only $E_-^2$ can become negative, the
criterion for dynamic stability is
\begin{equation}
\label{dynamic stability condition}
E_A^2E_B^2-4\epsilon_A \epsilon_B g_{AB}^2 > 0.
\end{equation}
In view of Eqs.~(\ref{Char_Equ1}) and (\ref{Char_Equ2}), this is
equivalent to the condition
\beq
{\rm det}(\sigma_z\mathcal{M}) = {\rm det}(\mathcal{M})
> 0,
\eeq
since ${\rm det}(\sigma_z) =1$.
Using the definition of $E_s^2$ in Eq.~(\ref{Bog_freq}) and defining
\beq
\gamma_{ss'} \equiv \frac{2\sqrt{M_sM_{s'}}R^2}{\hbar^2}
g_{ss'},
\label{gamma}
\eeq
Eq.~(\ref{dynamic stability condition}) becomes
\begin{equation}
\label{dynamic stability condition final}
\left (\gamma_{AA}+ \frac{1}{2}m^2\right ) \left (\gamma_{BB}+
\frac{1}{2} m^2\right ) > 
\gamma_{AB}^2.
\end{equation}
For repulsive interactions, this inequality is satisfied for all
$m$ if it is satisfied for $m=1$. This limiting case gives the
condition
\begin{equation}
\left (\gamma_{AA}+ \frac{1}{2}\right ) \left (\gamma_{BB}+
\frac{1}{2}\right ) > 
\gamma_{AB}^2.
\end{equation}
A criterion of this form was obtained in~\cite{Smyrnakis09} 
for $M_A=M_B$ but is also seen to be valid for $M_A \ne M_B$
with the definition of $\gamma_{ss'}$ given in Eq.~(\ref{gamma}). 

To complete our discussion of the Bogoliubov excitations we
present the results for the Bogoliubov amplitudes. It is
straightforward to show that
Eqs.~(\ref{Bogo_equ}) and (\ref{Normalization}) lead to
\bea
\big (\tilde{u}^{(\pm)}_{s,m}\big )^2&=&\frac{(E_\pm+E_s)^2
(E^2_\pm-E^2_{\bar s})}{4E_\pm E_s(2E^2_\pm- E^2_A-E^2_B)} 
\label{interm_Bog_amp_u}
\\
\big ( \tilde{v}^{(\pm)}_{s,m}\big )^2&=&\frac{(E_\pm-E_s)^2
(E^2_\pm-E^2_{\bar s})}{4E_\pm E_s(2E^2_\pm- E^2_A-E^2_B)}.
\label{interm_Bog_amp_v}
\eea
where $\bar s$ denotes the species complementary to $s$.
Finally, the relation of the original creation and annihilation
operators to the Bogoliubov quasiparticle operators is defined via
\bea
&&\hat a_{s,m}=u^{(+)}_{s,m}\hat 
\beta_{+,m}-v^{(+)}_{s,m}\hat\beta^\dag_{+,-m}+u^{(-)}_{s,m}\hat 
\beta_{-,m}-v^{(-)}_{s,m}\hat \beta^\dag_{-,-m} \nn \\
&&\hat a_{s,-m}=u^{(+)}_{s,m}\hat 
\beta_{+,-m}-v^{(+)}_{s,m}\hat\beta^\dag_{+,m}+u^{(-)}_{s,m}\hat 
\beta_{-,-m}-v^{(-)}_{s,m}\hat \beta^\dag_{-,m}.
\eea
These amplitudes can be obtained from Eq.~(\ref{bare_Bog_amp}) and 
Eqs.~(\ref{interm_Bog_amp_u}) and (\ref{interm_Bog_amp_v}) 
with the result
\bea
\label{Bog_amp}
\big ( {u^{(\pm)}_{s,m}} \big
)^2=\frac{(E_\pm+\epsilon_s)^2(E^2_\pm-E^2_{\bar s})
}{4E_\pm \epsilon_s(2E^2_\pm-E^2_A-E^2_B)} \\
\big ({v^{(\pm)}_{s,m}}\big )^2=\frac{(E_\pm-\epsilon_s)^2(E^2_\pm-E^2_{\bar
s}) }{4E_\pm \epsilon_s(2E^2_\pm-E^2_A-E^2_B)}.
\eea
It can be shown that these expressions are equivalent to those
given in Ref.~\cite{Tommasini03} in the one-dimensional limit.
The amplitudes can be used to evaluate the mode density fluctuations
$\delta n_{s,m}^{(\pm)}(\theta)$ of each species. We find that the
$A$ and $B$ density fluctuations are {\it in-phase} for the
(+) mode and {\it out-of-phase} for the (--) mode.

We now make use of these results in Eq.~(\ref{lc3})
to investigate the possibility of persistent currents at the 
angular momenta $L_n=n N\hbar$ for the equal-mass system.
The lower of the two branches in Eq.~(\ref{Bog_spec}) is the branch
relevant to determining the stability of the current. For
$M_A=M_B=M$, the energy of this branch reads
\beq
E_-(m)=\frac{\hbar^2}{2MR^2}\sqrt {m^4+m^2\left 
(\gamma_{AA}+\gamma_{BB}-\sqrt{(\gamma_{AA}-\gamma_{BB})^2+4\gamma_{AB}^
2}\right )}.
\label{lower_branch}
\eeq
According to Eq. (\ref{lc3}), the stability of persistent
currents at $L_n$ requires
\beq
\frac{n\hbar}{MR^2}<\frac{\hbar}{2MR^2}\sqrt 
{1+\gamma_{AA}+\gamma_{BB}-\sqrt{(\gamma_{AA}-\gamma_{BB})^2+4\gamma_{AB
}^2}}.
\eeq
This inequality is satisfied if the following two inequalities
\bea
\left (\gamma_{AA}-\frac{4n^2-1}{2}\right )\left 
(\gamma_{BB}-\frac{4n^2-1}{2}\right )&>&\gamma^2_{AB} 
\label{cri_g1} \\
\gamma_{AA}+\gamma_{BB}&>&4n^2-1 ,
\label{cri_g2}
\eea
are simultaneously satisfied. In the limit $\gamma_{AB}=0$, we
have two independent components and we observe that the
inequalities are satisfied if $\gamma_{\rm min} = {\rm
min}(\gamma_{AA},\gamma_{BB})$ satisfies
\beq
\gamma_{\rm min} > \frac{1}{2}(2n+1)(2n-1).
\label{gamma_crit}
\eeq
For $n =1$ this gives the critical interaction strength
$\gamma_{\rm cr} = 3/2$ which is the value quoted in
Ref.~\cite{Smyrnakis09}.

For the two-species system with equal masses, the inequalities
in Eqs.~(\ref{cri_g1}) and (\ref{cri_g2}) can usually be satisfied for
suitable choices of the interaction parameters, implying the
possible stability of persistent currents at any $L_n$.
The only exception occurs when 
\beq
\gamma_{AA}\gamma_{BB}=\gamma^2_{AB},
\eeq
or equivalently 
\beq
U_{AA}U_{BB}=U_{AB}^2.
\label{U_ij_equality}
\eeq
In this case, the coefficient of $m^2$ in Eq.~(\ref{lower_branch})
vanishes and the lower branch has a free particle dispersion
which destabilizes persistent currents for any value of $n$.
This conclusion was arrived at earlier by Smyrnakis {\it et
al.}~\cite{Smyrnakis09} for the special case $U_{AA} = U_{BB} = U_{AB}$; 
we see here how it follows from the Landau criterion for the
more general relation in Eq.~(\ref{U_ij_equality}).
However, this does not preclude the possibility of persistent
currents at non-integral values of angular momentum per particle. 
In the next
section we reconsider the problem from the point of view of
mean-field theory, following closely the work of Smyrnakis {\it
et al.}~\cite{Smyrnakis09}

\section{Persistent currents at non-integer angular momentum per 
particle: mean-field theory}
\label{GP-Theory}

The analysis in this section is based on the mean-field
Gross-Pitaevksii energy functional for the two-component system
in the ring geometry:
\begin{eqnarray}
\label{4.2_energy_1}
&&\hspace{-9mm}E[\psi_A, \psi_B] = \int_{0}^{2\pi} d\theta \left( 
\frac{N_A\hbar^2}{2M_AR^2} \left| \frac{d \psi_A}{d \theta} \right|^2 + 
\frac{N_B\hbar^2}{2M_BR^2} \left| \frac{d \psi_B}{d \theta} \right|^2 
\right) \nonumber \\
&& \hspace{18mm}+\hspace{1mm} 
\frac{1}{2}U_{AA}N_A^2\int_{0}^{2\pi} d\theta |\psi_A|^4 + 
\frac{1}{2}U_{BB}N_B^2\int_{0}^{2\pi} d\theta |\psi_B|^4 
+ U_{AB}N_A N_B\int_{0}^{2\pi} d\theta 
|\psi_A|^2 |\psi_B|^2.
\end{eqnarray}
Here the condensate wave functions $\psi_A$ and $\psi_B$ are
normalized as
\begin{equation}
\int_0^{2\pi} d\theta |\psi_A(\theta)|^2 =  \int_0^{2\pi}
d\theta |\psi_B(\theta)|^2 = 1.
\end{equation}
As discussed in the previous section, Bloch's argument allows for 
persistent currents at integral values of $l=L/N\hbar$ when 
$M_A=M_B=M$ except when Eq.~(\ref{U_ij_equality}) is true. Here,
following Smyrnakis {\it et al.}~\cite{Smyrnakis09}, 
we consider the special case
$U_{AA}=U_{BB}=U_{AB}=U$. In units of the energy $N\hbar^2/(2MR^2)$, 
Eq.~(\ref{4.2_energy_1}) becomes
\begin{eqnarray}
\label{4.2_energy_2}
&&\hspace{-9mm}\bar{E}[\psi_A, \psi_B] =  \int_{0}^{2\pi} d\theta 
\left( x_A \left| \frac{d \psi_A}{d \theta} \right|^2 + x_B \left| 
\frac{d \psi_B}{d \theta} \right|^2 \right) \nonumber \\
&& \hspace{18mm}+  \hspace{1mm} x_A^2\pi\gamma \int_{0}^{2\pi} d\theta 
|\psi_A|^4 + x_B^2\pi\gamma\int_{0}^{2\pi} d\theta |\psi_B|^4 
+  \hspace{1mm}2x_Ax_B\pi\gamma \int_{0}^{2\pi}d\theta 
|\psi_A|^2 |\psi_B|^2,
\label{GP_functional}
\end{eqnarray}
where $x_A=N_A/N$, $x_B=N_B/N$ are the relative fractions of the two 
species in the system and $\gamma \equiv NMR^2U/\pi\hbar^2$ is
a dimensionless interaction parameter. For definiteness, we 
take $N_A > N_B$.

The objective is to minimize the energy functional in
Eq.~(\ref{GP_functional}) with the constraint that 
that the average value of the total angular
momentum has a fixed value $L\equiv lN\hbar$. This is achieved
by expanding the condensate wave functions as
\begin{eqnarray}
\label{4.2_wfA}
&&\psi_A(\theta) = \sum\limits_{m}c_m\phi_m(\theta) \\
\label{4.2_wfB}
&&\psi_B(\theta) = \sum\limits_{m}d_m\phi_m(\theta),
\end{eqnarray}
where the basis functions $\phi_m(\theta)$ are given in
Eq.~(\ref{s_p_basis}).
The normalization of the wave functions requires
\begin{equation}
\label{4.2_normalization}
\sum\limits_{m}|c_m|^2=1,\hspace{7mm} \sum\limits_{m}|d_m|^2=1.
\end{equation}
Such a superposition implies
that the wave functions are in general nonuniform around the
ring. In addition, 
the expansion coefficients $c_m$ and $d_m$ must satisfy
the angular momentum constraint
\begin{equation}
l = x_Al_A+x_Bl_B \equiv x_A \sum\limits_{m}m|c_m|^2 +
x_B \sum\limits_{m}m|d_m|^2.
\label{ang_mom_constraint}
\end{equation}
$l_A$  ($l_B$) represents the average angular momentum in units 
of $\hbar$ of an $A$ ($B$)-species particle. The minimization
of the energy with respect to the expansion coefficients in
Eqs.~(\ref{4.2_wfA}) and (\ref{4.2_wfB})
was first considered by Smyrnakis {\it et al.}~\cite{Smyrnakis09}. 
It will be clear from
the following that much of our analysis closely follows theirs.
However, we have expanded on their discussion in order to obtain a
number of results that are not given explicitly in their paper.

Substituting the wave functions in Eqs.~(\ref{4.2_wfA}) and 
(\ref{4.2_wfB}) into Eq.~(\ref{4.2_energy_2}), we obtain
\begin{eqnarray}
&&\bar{E}_0(l) = 
x_A\sum\limits_{m}m^2|c_m(l)|^2+x_B\sum\limits_{m}m^2|d_m(l)|^2+x_A^2\pi
\gamma\int_0^{2\pi}d\theta\Big|\sum\limits_{m}c_m(l)\phi_m(\theta) 
\Big|^4 \nonumber \cr
&&\hspace{15mm}+\hspace{1mm}x_B^2\pi\gamma\int_0^{2\pi}d\theta\Big|\sum
\limits_{m}d_m(l)\phi_m(\theta) \Big|^4 
+ 2x_Ax_B\pi\gamma\int_0^{2\pi}\Big|\sum\limits_{m}c_m(l)\phi_
m(\theta) \Big|^2\Big|\sum\limits_{m}d_m(l)\phi_m(\theta)
\Big|^2 \nonumber \cr
&&\hspace{9mm} \equiv l^2 + \bar e_0(l).
\label{4.2_energy_3}
\end{eqnarray}
According to Bloch's argument, $\bar e_0(l)$ should exhibit the
periodicity $\bar e_0(l+n) = \bar e_0(l)$ where $n$
is an integer. This periodicity is ensured if the expansion
coefficients satisfy the periodicity conditions
\begin{equation}
c_{m+n}(l+n)=c_m(l), \quad d_{m+n}(l+n)=d_m(l).
\label{periodicity}
\end{equation}
The fact that $\bar E_0(l)$ must remain unchanged when the
wave functions $\psi_\alpha^*(\theta)$ 
with angular momenta $-l_\alpha$ are used to evaluate the energy
functional leads to the relations
\begin{equation}
c_m(-l)=c_{-m}^*(l), \quad d_m(-l)=d_{-m}^*(l).
\label{reflection}
\end{equation}
These two conditions are the mean-field counterparts of
Eqs.~(\ref{e_period}) and (\ref{e_inversion}).

The function $\bar e_0(l)$ is the central quantity determining the
possibility of persistent currents and its detailed evaluation
is taken up next.
To begin, we consider wave functions $\psi_A$ and $\psi_B$ containing 
only two components, that is,
\begin{eqnarray}
\label{4.2_wfA_01}
&&\psi_A = c_0\phi_0+c_1\phi_1 \\
\label{4.2_wfB_01}
&&\psi_B = d_0\phi_0+d_1\phi_1.
\end{eqnarray}
The coefficients $c_m$ and $d_m$ are normalized according to 
Eq.~(\ref{4.2_normalization}) and the angular momentum constraint
becomes
\begin{equation}
\label{4.2_ang_mom_01_full}
x_A|c_1|^2+x_B|d_1|^2=l.
\end{equation}
Expressing the complex coefficients in the form 
\begin{eqnarray}
\label{4.2_complex_form_coeff_c}
&&c_m = |c_m|e^{i\alpha_m}\\
\label{4.2_complex_form_coeff_d}
&&d_m = |d_m|e^{i\beta_m},
\end{eqnarray}
the GP energy becomes
\begin{equation}
\label{4.2_energy_5}
\bar{E}_0(l) = 
l+\frac{\gamma}{2}+\gamma\left(x_A^2|c_0|^2|c_1|^2 
+x_B^2|d_0|^2|d_1|^2 
+2x_Ax_B|c_0||c_1||d_0||d_1| \cos\chi\right )
\end{equation}
where $\chi = \alpha_0 -\alpha_1-\beta_0+\beta_1$.
The choice of $\chi$ which minimizes $\bar{E}_0(l)$ is $\pi$ 
and we then have
\begin{equation}
\bar{E}_0(l) = 
l+\frac{\gamma}{2}+\gamma\left(x_A|c_0||c_1| 
-x_B|d_0||d_1|\right )^2.
\end{equation}
The lowest possible value of this energy is~\cite{Smyrnakis09}
\begin{equation}
\label{4.2. energy simplified}
\bar{E}_0(l)=l+\gamma/2,
\end{equation}
which occurs for 
\begin{equation}
\label{4.2. simplification}
x_A|c_0||c_1| = x_B|d_0||d_1|.
\end{equation}
This relation, together with the normalization and angular
momentum constraints, yields the coefficients
\begin{eqnarray}
\label{4.2_coeff_csq}
&&|c_0|^2 = \frac{(x_A -l)(1-l)}{x_A(1-2l)}, \quad 
|c_1|^2 = \frac{l(x_B -l)}{x_A(1-2l)} \\
\label{4.2_coeff_dsq}
&&|d_0|^2 = \frac{(x_B -l)(1-l)}{x_B(1-2l)}, \quad
d_1|^2 = \frac{l(x_A -l)}{x_B(1-2l)}.
\end{eqnarray}
These quantities are positive 
provided $l$ is in the range $0 \leq l \leq x_B$ 
or $x_A\leq l \leq 1$. Assuming the validity of 
Eq.~(\ref{4.2. energy simplified}) for $l$ in these ranges, we see
that $\bar E_0(l)$ does not have a local minimum
at $l =1$. Thus, persistent currents are not possible at $l=1$,
and by virtue of the periodicity of $\bar e_0(l)$, at all integral 
values of $l$. These conclusions are consistent with our
earlier discussion based on the Landau criterion; the validity of
Eq.~(\ref{U_ij_equality}) implies the existence of
particle-like excitations and the absence of persistent currents
at integral values of $l$.
 
Although Eq.~(\ref{4.2. energy simplified}) was obtained for the
simplest possible variational wave function, it in fact is
exact when $l$ is restricted to the above
ranges~\cite{Smyrnakis09}.
To show this,
we consider normalized wave functions of the form 
\begin{equation}
\label{4.2_wf_perturbed}
\tilde \psi_A(\theta) = \psi_A(\theta)+\delta\psi_A, \hspace{1cm} 
\tilde \psi_B(\theta) = \psi_B(\theta)+\delta\psi_B,
\end{equation}
where $\psi_A$ and $\psi_B$ are defined by Eqs.~(\ref{4.2_wfA_01}) and
(\ref{4.2_wfB_01}) with the coefficients given
in Eqs.~(\ref{4.2_coeff_csq})-(\ref{4.2_coeff_dsq}).
If the deviations are expressed in the form
\begin{equation}
\label{4.2_wf_deviations}
\delta\psi_A = \sum\limits_{m}\delta c_m\phi_m, \hspace{1cm} 
\delta\psi_B = \sum\limits_{m}\delta d_m\phi_m,
\end{equation}
the angular momentum constraint in Eq.~(\ref{ang_mom_constraint})
leads to
\begin{equation}
\label{4.2_deviations_coeff_cancel}
x_Ac_1(\delta c_1+\delta c_1^*) + x_Bd_1(\delta d_1+\delta d_1^*)= 
-x_A\sum\limits_{m}m|\delta c_m|^2-x_B\sum\limits_{m}m|\delta d_m|^2.
\end{equation}
We next observe that the density $n_0(\theta)
=N_A|\psi_A|^2+N_B|\psi_B|^2$ is in fact uniform, that is,
$n_0(\theta) = N/(2\pi)$.
Using these results, the energy is found to be given by
\begin{equation}
\bar{E}[\tilde\psi_A,\tilde \psi_B] = \bar E_0(l) 
+ x_A\sum_{m}(m^2-m)|\delta 
c_m|^2+x_B\sum\limits_{m}(m^2-m)|\delta d_m|^2 
+ \frac{\pi\gamma}{N^2}\int_0^{2\pi} 
d\theta |\delta n(\theta)|^2,
\end{equation}
where $\delta n(\theta) = n(\theta)-n_0$. We thus see that $\bar
E[\tilde\psi_A,\tilde \psi_B] > \bar E_0(l)$, implying that the 
state defined by Eqs.~(\ref{4.2_wfA_01}) and (\ref{4.2_wfB_01})
is indeed the ground state of the
system for the assumed ranges of the angular momentum. It should
be noted that this result depends crucially on the assumption of
equal interaction parameters between all components. The weaker
condition in Eq.~(\ref{U_ij_equality}) still precludes the possibility
of persistent currents at integral values of $l$, but the energy
does not have the simple form shown in Eq.~(\ref{4.2. energy
simplified}).

We next analyze the energy for $x_B\leq l\leq x_A$. In 
particular we consider the situation when $l$ is close to $x_A$, that 
is $l-x_A=-\varepsilon$, where $\varepsilon$ is a small positive
quantity. For $l=x_A$ we see from 
Eqs.~(\ref{4.2_coeff_csq})-(\ref{4.2_coeff_dsq}) that 
\begin{eqnarray}
\label{4.2._coeff small c}
&&|c_0|^2=0, \hspace{1cm} |c_1|^2=1\\
\label{4.2._coeff small d}
&&|d_0|^2=1, \hspace{1cm} |d_1|^2=0.
\end{eqnarray}
As $\varepsilon$ increases from zero, we therefore expect deviations 
from these limiting values and additional components in the expansion 
of the $\psi_A$ and $\psi_B$ wave functions. To be specific, we 
consider the three-component wave functions  
\begin{eqnarray}
\label{4.2 3comp wfA}
&&\psi_A=c_0\phi_0+c_1\phi_1+c_2\phi_2 \\
\label{4.2 3comp wfB}
&&\psi_B=d_{-1}\phi_{-1}+d_0\phi_0+d_1\phi_1.
\end{eqnarray}
We anticipate that $|c_0|^2$, $|c_2|^2$, $|d_{-1}|^2$ and $|d_1|^2$ are 
all of order $\varepsilon$. With this assumption, 
the energy to first order in $\varepsilon$ is found to be
\begin{eqnarray}
\label{E_3_component}
&&\hspace{-9mm}\bar{E}_0(l) =  x_A\left(|c_1|^2 + 4|c_2|^2 \right) + 
x_B\left(|d_{-1}|^2 + |d_1|^2 \right) +\frac{\gamma}{2}\\ \nonumber
&&+\hspace{1mm}x_A^2\gamma\left(|c_0|^2+|c_2|^2 + 
2|c_0||c_2|\cos\chi_1 \right)
+x_B^2\gamma\left(|d_{-1}|^2+|d_1|^2 + 
2|d_{-1}||d_1|\cos\chi_2 \right) \\ \nonumber
&&+\hspace{1mm}2x_Ax_B\gamma\left[ 
\frac{1}{2}+|c_0||d_{-1}|\cos\chi_3 + 
|c_0||d_1|\cos(\chi_3-\chi_2)
+|c_2||d_{-1}|\cos(\chi_3-\chi_1)
+|c_2||d_1|\cos(\chi_3-\chi_1-\chi_2)\right],
\end{eqnarray}
where we have defined the phase angles
$\chi_1 = \alpha_0-2\alpha_1+\alpha_2$, $\chi_2 =
\beta_{-1}-2\beta_0+\beta_1$ and $\chi_3 =
\alpha_0-\alpha_1-\beta_{-1}+\beta_0$. This energy is an
extremum with respect to the phase angles if they are all 0 or
$\pi$. If we choose them arbitrarily to be 0, we obtain
\begin{equation}
\label{4.2. E new 2}
\bar{E}_0(l) \simeq x_A\left(|c_1|^2 + 4|c_2|^2 \right) + 
x_B\left(|d_{-1}|^2 + |d_1|^2 \right) +\frac{\gamma}{2}
+\gamma\left[ x_A(|c_0| + |c_2|)+ x_B(|d_{-1}| + |d_1|) \right]^2,
\end{equation}
which must now be minimized with respect to the coefficients $|c_0|$, 
$|c_2|$, $|d_{-1}|$ and $|d_1|$ subject to the angular momemtum
constraint
\begin{equation}
\label{4.2. ang mom new}
l =x_A - \varepsilon = x_A\left(|c_1|^2 + 2|c_2|^2 \right) + 
x_B\left(|d_{1}|^2 -|d_{-1}|^2 \right)= x_A\left(1-|c_0|^2 +
|c_2|^2 \right) + x_B\left(|d_{1}|^2 -|d_{-1}|^2 \right).
\end{equation}
If this minimization in the end leads 
to coefficients that are negative, the phases have to be
adjusted accordingly to yield coefficients with positive
values. As we shall see, this will indeed be necessary.

Using Eq.~(\ref{4.2. ang mom new}) to eliminate $|c_1|$ from
Eq.~(\ref{4.2. E new 2}), and introducing a Lagrange multiplier
$\lambda$ to account for the angular momentum constraint, the
functional to be minimized is
\begin{eqnarray}
\label{4.2. lagrange 00}
&&\hspace{-24mm}F(|c_0|, |c_2|, |d_{-1}|, |d_1|) = 
2x_A|c_2|^2 + 2x_B|d_{-1}|^2
+ \gamma\Big[ x_A\left( |c_0| + |c_2| 
\right) + x_B\left( |d_{-1}| + |d_1| \right) \Big]^2 \nonumber \\ 
&&\hspace{27.5mm}+ \hspace{1mm}\lambda \Big[ x_A\left(1- |c_0|^2 + 
|c_2|^2 \right) + x_B\left( -|d_{-1}|^2 +|d_1|^2 \right) \Big],
\end{eqnarray}
where the variations of the coefficients are now unconstrained.
This variation leads to the results
\begin{equation}
\label{4.2. lambda cd1}
\left |\frac{c_2}{c_0}\right | =
-\frac{\lambda}{\lambda+2},\quad
\left | \frac{d_{1}}{c_0}\right | = -1,\quad
\left |\frac{d_{-1}}{c_0}\right |=\frac{\lambda}{\lambda-2},
\end{equation}
where the Lagrange multiplier $\lambda$ is the solution of the cubic
equation~\cite{Smyrnakis09}
\begin{equation}
\label{4.2 f of lambda}
f(\lambda)\equiv \lambda(\lambda^2-4)-2\gamma\lambda 
+4\gamma(x_A-x_B)=0.
\end{equation}
The roots of this equation are to be determined for 
$\gamma>0$ and $0\leq (x_A-x_B)\leq 1$. 

\begin{figure*}[!ht]
\centering \scalebox{0.4}
{\includegraphics[50,200][575,575]{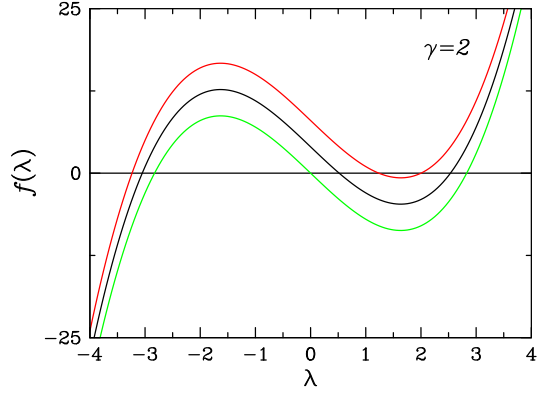}}
\caption{Plot of the cubic $f(\lambda)$ vs. $\lambda$. The
curves from bottom to top correspond to 
$x_A-x_B = 0$, 0.5 and 1.0. The interaction parameter is $\gamma
= 2$.
}
\label{fig:cubic_gamma_2} 
\end{figure*}
\begin{figure*}
\centering \scalebox{0.4}
{\includegraphics[50,200][575,575]{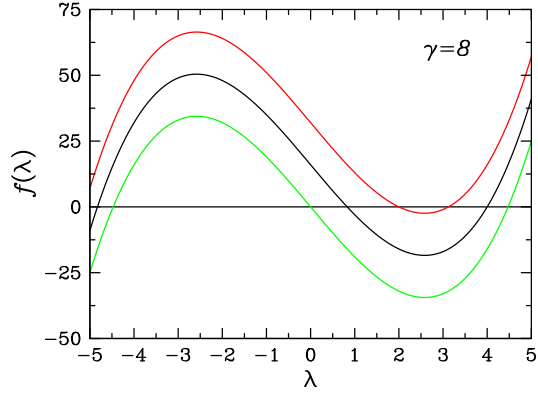}}
\caption{As for Fig.~\ref{fig:cubic_gamma_2} but for an interaction
parameter of $\gamma = 8$.
}
\label{fig:cubic_gamma_8}
\end{figure*}
In Fig.~\ref{fig:cubic_gamma_2}, $f(\lambda)$ is plotted for 
$(x_A-x_B)=0$, 0.5 and 1 and for $\gamma = 2$;
Fig.\ref{fig:cubic_gamma_8} is a similar plot for $\gamma = 8$. For 
$(x_A-x_B)=0$, $f(\lambda)=\lambda(\lambda^2-4-2\gamma)$, which has the 
roots $\lambda=0$ and $\lambda=\pm\sqrt{4+2\gamma}$. For $(x_A-x_B)=1$, 
$f(\lambda)=(\lambda-2)[\lambda(\lambda+2)-2\gamma]$, which has the 
roots $\lambda=2$ and $\lambda=-1\pm\sqrt{1+2\gamma}$. The
latter two values 
are the Lagrange multipliers in the single-species limit as obtained 
from the minimization of Eq.~(\ref{4.2. lagrange 00}) for $x_B = 0$.
Since the term $4\gamma(x_A-x_B)$ in $f(\lambda)$ simply 
shifts the curves in Figs.~\ref{fig:cubic_gamma_2} and
\ref{fig:cubic_gamma_8} vertically, it is clear that there are 
always three real roots for the physical range of $(x_A-x_B)$ values.
For any positive value of $\gamma$,
one root is always less than $-2$, a second lies in the range $0\le
\lambda \le 2$ (more precisely in the range $0\le \lambda \le
2(x_A-x_B)$) and a third in the range $\lambda \ge 2$.
Substituting the coefficients given in Eq.~(\ref{4.2. lambda cd1}) 
into Eq.~(\ref{4.2. ang mom
new}) we find
\begin{equation}
\label{4.2. c0 as eps}
|c_0|^2
=\frac{\varepsilon(\lambda^2-4)^2}{4[x_A(\lambda+1)
(\lambda-2)^2+x_B(\lambda-1)(\lambda+2)^2]}.
\end{equation}
It is clear from this expression that the $\lambda<-2$ root makes 
$|c_0|^2$ negative. This root is therefore physically inadmissible 
and only the positive $\lambda$ roots are relevant.
Eq.~(\ref{4.2. c0 as eps}) together with Eq.~(\ref{4.2. lambda cd1})
can be used in Eq.~(\ref{4.2. E new 2}) to evaluate the energy. One
finds the remarkably simple result
\begin{equation}
\label{4.2. E new 5 simple}
\bar{E}_0(l)-\frac{\gamma}{2} =
x_A-\varepsilon+\lambda\varepsilon = x_A+(l-x_A)(1-\lambda).
\end{equation}
We now see that the smaller of the two positive $\lambda$ roots gives 
the lowest possible energy. This thus identifies the root in the range 
$0<\lambda <2$ as the one that is physically
relevant~\cite{Smyrnakis09}. For $\lambda$ in this range we
observe that the 
ratios in Eq.~(\ref{4.2. lambda cd1}) are negative, indicating that
the phases in Eq.~(\ref{E_3_component}) were chosen incorrectly. The
proper phases are $\chi_1=\pi$, $\chi_2=0$ and $\chi_3=\pi$.

The criterion for the existence of persistent currents at
$l=x_A$ used in Ref.~\cite{Smyrnakis09} is that the slope 
of $\bar E_0(l)$ in Eq.~(\ref{4.2. E new 5 simple}) at
$l=x_A^-$ is negative, i.e., $\lambda > 1$. The critical condition
is thus $\lambda = 1$, which from Eq.~(\ref{4.2 f of lambda}) gives
the critical interaction strength~\cite{Smyrnakis09}
\begin{equation}
\label{4.2. gamma cr}
\gamma_{cr} =\frac{3}{4(x_A-x_B)-2} = \frac{3}{2(4x_A-3)}.
\end{equation}
In the $x_A =1$ limit this reduces to $\gamma_{cr} = 3/2$ which
is the value obtained at $l=1$ for the single-species system.
To obtain the critical coupling at $l=x_A+n-1$, where $n =
1$, 2,.., we write $\bar E_0(l) = l^2 +\bar e_0(l)$ and
use the fact that $\bar e_0(l)$ is periodic. The slope at
$l=(x_A+n-1)^-$ is thus found to be
\begin{equation}
\label{4.2. denergy prime}
\left. \frac{d\bar{E}_0(l)}{dl}\right|_{l=(x_A+n-1)^-} = 
2n-1-\lambda.
\end{equation}
If the root in the range $0<\lambda<2(x_A-x_B)$ is used, 
the slope cannot be zero for any $n>1$. This is the basis of
the claim made in Ref.~\cite{Smyrnakis09} that 
persistent currents are not possible for $l>1$; seemingly,
an arbitrarily small amount of
the minority component $B$ has a profound effect on the
possibility of persistent currents. For the single-species case,
the energy is given by Eq.~(\ref{4.2. E new 5
simple}) with $x_A=1$, but the appropriate value of $\lambda$ is
$\lambda =-1+\sqrt{1+2\gamma}$, which is not bounded as a
function of $\gamma$. Using this value in Eq.~(\ref{4.2. denergy
prime}), one finds that persistent currents are
possible for all $n$ in this case, with a critical interaction
strength of $\gamma_{cr,n}={(2n+1)(2n-1)}/{2}$. This is the
result found earlier (Eq.~(\ref{gamma_crit})) 
using the Landau criterion. This comparison indicates an
inconsistency. On the one hand,
Eq.~(\ref{4.2. denergy prime}) does allow for
persistent currents for $l>1$ in the single-species limit if the
appropriate value of $\lambda$ is used. However, the two-species
analysis requires that the root in the range $0<\lambda<2$
be used, which precludes the possibility of persistent
currents for $l>1$ for any nonzero value of $x_B$. Since the
energy functional in Eq.~(\ref{4.2_energy_2}) reduces to the
single-species case when $x_B=0$, it would appear that taking
the $x_B\to 0$ limit of the two-species analysis is problematic.

In order to explain this discrepancy it is
useful to examine the behaviour of the coefficients in
Eqs.~(\ref{4.2. lambda cd1}) and (\ref{4.2. c0 as eps}) in the $x_A
\to 1$ limit in more detail. These coefficients are determined
by the root $\lambda$ that lies in the range
$0 \le \lambda < 2$. If $\gamma < 4$, the limiting value of this
root for $x_A\to 1$ is $\lambda = -1 +\sqrt{1+2\gamma}$.
This is the $\lambda$ value for the single-species
case. Thus for this range of $\gamma$, one recovers the
single-species values for all the coefficients. However, for
$\gamma > 4$, the root in the range $0 \le \lambda < 2$ has the
limiting value of 2 which is less than the $\lambda =
-1+\sqrt{1+2\gamma}$ root. The limiting values of the
coefficients do not correspond to the single-species values in
this case.

The distinction between $\gamma < 4$ and $\gamma>4$ is revealed
more clearly by plotting the coefficients in these two cases as
a function of $x_B$. We observe that
the angular momenta carried by each of the species is given by
\begin{equation}
\label{4.2. la}
l_A = x_A\left( |c_1|^2+2|c_2|^2 \right) = x_A+x_A\left( 
|c_2|^2-|c_0|^2\right)
\end{equation}
and
\begin{equation}
\label{4.2. lb}
l_B = x_B\left( -|d_{-1}|^2+|d_1|^2 \right).
\end{equation}
The change in angular momentum as $l$ is reduced from $x_A$ 
is associated with the transfer of weight from
one angular momentum component to another. For example, for the
$A$ species, the transfer takes place from the $m=1$ state
to the $m=0$ or $m=2$ states, with respectively, a decrease or
increase in angular momentum. For the $B$ species, the transfer
takes place from the $m=0$ state to the $m=-1$ and $m=1$ states.
Of interest is the relative magnitude of the
angular momentum change $\Delta l_{s,m}$
that is associated with each angular
momentum component. We therefore define the ratios $\Delta
l_{s,m}/(-\varepsilon)$ where for example, $\Delta
l_{A,0}/(-\varepsilon) = (-x_A|c_0|^2)/(-\varepsilon)$.
These
ratios represent the fraction of the angular momentum
change $-\varepsilon = l-x_A$
attributable to each of the angular momentum components.
\begin{figure*}[t]
\centering \scalebox{0.4}
{\includegraphics[50,200][575,575]{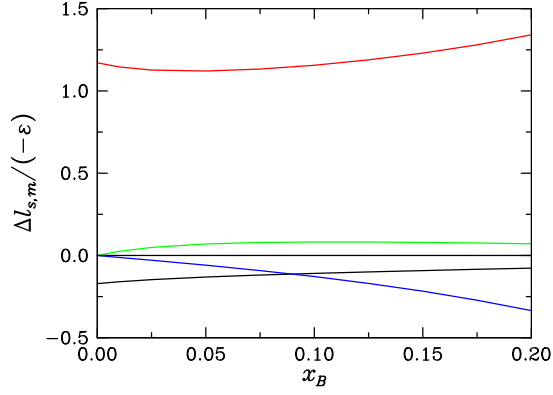}}
\caption{The angular momentum change carried by each of the wave
function components relative to the total angular momentum
change of $-\varepsilon$ as a function of $x_B$: red ($\Delta
l_{A,0}$), black ($\Delta l_{A,2}$), green ($\Delta l_{B,-1}$),
blue ($\Delta l_{B,1}$). The interaction parameter is $\gamma
=2$.
}
\label{fig:coeff_gamma_2} 
\end{figure*}
\begin{figure*}[!ht]
\centering \scalebox{0.4}
{\includegraphics[50,200][575,575]{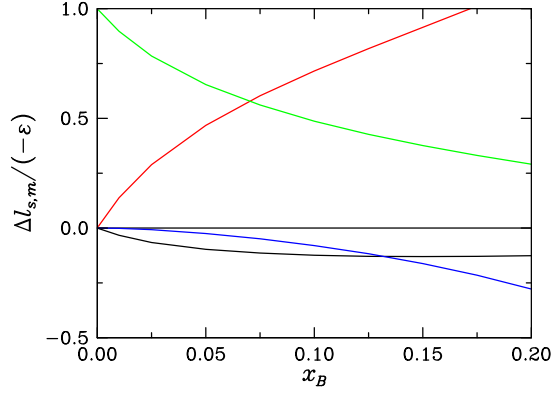}}
\caption{As in Fig.~\ref{fig:coeff_gamma_2} but for $\gamma =8$.
}
\label{fig:coeff_gamma_8} 
\end{figure*}
In Fig.~\ref{fig:coeff_gamma_2} we plot 
these ratios as a function of $x_B$ for $\gamma =2$;
Fig.~\ref{fig:coeff_gamma_8} gives similar plots for 
$\gamma =8$. For $\gamma =2 $, we see that species $B$ carries a
relatively small contribution of the angular momentum change.
This contribution vanishes in the 
$x_B\rightarrow 0$ limit and the situation reverts to that of
the single species which, as discussed above, is generally the case 
for $\gamma < 4$. The situation for $\gamma >4$, however, is
quite different. Fig.~\ref{fig:coeff_gamma_8} for $\gamma = 8$ 
shows that the angular 
momentum change is carried entirely by the $m=-1$ component of
the $B$ species in the 
$x_B\rightarrow 0$ limit. The reason for this surprising result is that 
the relevant $\lambda$ root approaches 2 for $x_B\rightarrow 0$
when $\gamma > 4$. Eq.~(\ref{4.2. c0 as eps}) then gives
$|c_0|^2 \simeq
(2-\lambda)^2\varepsilon/(4x_B)$ and from Eq.~(\ref{4.2. lambda
cd1}) we find
$|d_{-1}|^2\simeq \varepsilon/x_B$ for $x_B\rightarrow 0$, i.e.
$l_B = -\varepsilon$ in this limit. 
The divergence of $|d_{-1}|^2$ as 
$x_B\rightarrow 0$ is indicating that the result can only be valid for 
a decreasingly smaller range of $\varepsilon$ since the normalization 
$1=|d_{-1}|^2+|d_0|^2+|d_1|^2$ must be preserved.
In other words, the energy
$\bar E_0(l)$, as given by
Eq.~(\ref{4.2. E new 5 simple}), is meaningful in
an interval of $l$ of decreasing size as $x_B\to 0$. 

The above results call into question any
conclusion regarding the possibility of persistent currents at
higher angular momenta when $x_A$ approaches 1. In this limit, a
more global perspective regarding the behaviour of the energy as a
function of $l$ in the interval $x_B \le l \le x_A$ is required.
We now give a general argument for the possibility of persistent
currents at $l>1$ based on the assumption of continuity of the 
GP energy as a function of $x_B$.  To exhibit this dependence we
write $\bar E_0(l,x_B)$ and consider this function in the limit
of small $x_B$. In particular, we have $\bar E_0(n,x_B) = \bar
E_0^A(n)+\delta_1(x_B)$ and $\bar E_0(n-\Delta l,x_B) = \bar
E_0^A(n-\Delta l)+\delta_2(x_B)$ where $\bar E_0^A(l)= \bar
E_0(l,x_B=0)$ is the energy of the single-species system. 
The assumption of continuity implies that
$\delta_1(x_B)$ and $\delta_2(x_B)$ approach 0 as $x_B\to 0$.
We then have $\bar E_0(n-\Delta l,x_B) - \bar E_0(n,x_B) = \bar
E_0^A(n-\Delta l) - \bar E_0^A(n) + \delta_2(x_B) -
\delta_1(x_B)$. By choosing $\gamma > \gamma_{{\rm cr},n}$,
$\bar E_0^A(n-\Delta l) - \bar E_0^A(n)$ will have some fixed
{\it positive} value. Thus, we can say that $\bar E_0(n-\Delta
l,x_B) - \bar E_0(n,x_B) > 0$ for $x_B$ sufficiently small.
Since $\partial \bar E_0(l,x_B)/\partial l|_{l=n^-}< 0$, we
conclude that $\bar E_0(l,x_B)$ must have a local minimum
between $l=n-\Delta l$ and $l=n$. This argument can be used for
any $n$ and shows that persistent currents must be stable in the
vicinity of $l =n$ if $x_B$ is sufficiently small and $\gamma$
is sufficiently large.

Although it is difficult to evaluate $\bar E_0(l,x_B)$ for 
arbitrary $l$, the above general argument can be illustrated
quantitatively by evaluating the energy at $l = 1/2$.
To do so, it is sufficient to assume four-component 
wave functions of the form
\begin{eqnarray}
\label{4.2. 4 comp wf 2 species A}
&&\psi_A = c_{-1}\phi_{-1}+c_0\phi_0+c_1\phi_1+c_2\phi_2 \\
\label{4.2. 4 comp wf 2 species B}
&&\psi_B = d_{-1}\phi_{-1}+d_0\phi_0+d_1\phi_1+d_2\phi_2.
\end{eqnarray}
Substituting these wave functions into Eq.~(\ref{4.2_energy_2}), we have
\begin{eqnarray}
\label{4.2. 4 comp energy 1}
&&\hspace{-1.5cm}\bar{E}_0(l=1/2) = x_A\left( |c_{-1}|^2 + |c_1|^2 + 4|c_2|^2 
\right) + x_B\left( |d_{-1}|^2 + |d_1|^2 + 4|d_2|^2 \right)\nonumber\\
&&\hspace{-0.6cm}+\hspace{1mm}x_A^2\pi\gamma\int_0^{2\pi}d\theta 
|c_{-1}\phi_{-1}+c_0\phi_0+c_1\phi_1+c_2\phi_2|^4
+x_B^2\pi\gamma\int_0^{2\pi}d\theta 
|d_{-1}\phi_{-1}+d_0\phi_0+d_1\phi_1+d_2\phi_2|^4\nonumber \\
&&\hspace{-0.6cm}+\hspace{1mm}2x_Ax_B\pi\gamma\int_0^{2\pi}d\theta 
|c_{-1}\phi_{-1}+c_0\phi_0+c_1\phi_1+c_2\phi_2|^2 
|d_{-1}\phi_{-1}+d_0\phi_0+d_1\phi_1+d_2\phi_2|^2.
\end{eqnarray}
The periodicity and reflection
properties imply $c_0\left(\frac{1}{2}\right) =
c_1^*\left (\frac{1}{2}\right )$ and $c_{-1}\left
(\frac{1}{2}\right ) = c_2^*\left (\frac{1}{2}\right )$, with 
analogous relations for the $d_m$ amplitudes. These relations
reduce the number of variational parameters by half. We have in
particular
\begin{eqnarray}
\nonumber |c_0|=|c_1|\equiv x,&& \quad |c_{-1}|=|c_2|\equiv y \\
\nonumber \alpha_1=-\alpha_0,&& \quad \alpha_{-1}=-\alpha_2 \\
\nonumber |d_0|=|d_1|\equiv u,&& \quad |d_{-1}|=|d_2|\equiv v \\
\chi_1=-\chi_0, && \quad \chi_{-1}=-\chi_2.
\end{eqnarray}
Using these definitions, the normalization constraints reduce to
\begin{equation}
\label{4.2.2. xy norm cons}
x^2+y^2 = \frac{1}{2},\quad u^2+v^2 = \frac{1}{2}.
\end{equation}
Furthermore, the angular momentum of each species is given by
\begin{eqnarray}
\label{4.2. 4 comp la}
&&l_A = x_A(-|c_{-1}|^2 + |c_1|^2 + 2|c_2|^2) = x_A(x^2 + y^2)\\ 
\label{4.2. 4 comp lb}
&&l_B = x_B(-|d_{-1}|^2 + |d_1|^2 + 2|d_2|^2) = x_B(u^2 + v^2).
\end{eqnarray}
We thus see that normalization ensures that the total angular 
momentum has the required value of 1/2.

Using these results, the expression for the energy becomes
\begin{eqnarray}
\label{4.2.2 energy final}
&&\hspace{-1cm}\bar{E}_0(1/2) = \frac{1}{2}+ \frac{1}{2}\gamma
+ 4x_Ay^2+4x_Bv^2 \nonumber \\
&&\hspace{2.3mm}+ \hspace{1mm}x_A^2\gamma\Big[x^4 +y^4+ 
8x^2y^2+4x^3y\cos\beta\Big]
+x_B^2\gamma\Big[u^4+v^4+ 8u^2v^2+4u^3v\cos\xi\Big] \nonumber \\ 
&&\hspace{2.3mm}+\hspace{1mm}x_Ax_B\gamma\Big [8xyuv \left \{
\cos(\theta-\beta+\xi)+ \cos(2\theta-\beta+\xi)\right \}
+\hspace{1mm}4xyu^2\cos(\theta-\beta) + 
4x^2uv\cos(\theta+\xi) \Big . \nonumber \\
\Big.&&\hspace{2.5cm}+\hspace{1mm}2x^2u^2\cos\theta + 
2y^2v^2\cos(3\theta-2\beta+2\xi)\Big],
\end{eqnarray}
where we have defined the phase angles $\beta =
3\alpha_0+\alpha_2$, $\xi = 3\chi_0+\chi_2$ and $\theta =
2(\alpha_0-\chi_0)$.
We see that the energy depends on these three independent phases
and the two amplitudes $x$ and $u$. It clearly reduces 
to the single-species result in the $x_B\rightarrow 0$ limit.

For $x_A=1$, the energy 
is minimized for $\beta=\pi$ and a value of $x$ which is close
to $1/\sqrt{2}$.. We do not expect this conclusion 
to change when $x_A$ is close to, but not exactly equal to 1. For these 
values of $x_A$, the term in Eq.~(\ref{4.2.2 energy final}) 
proportional to 
$x_B^2$ is small and can be neglected. Setting $\beta=\pi$, the energy 
is approximately
\begin{eqnarray}
&&\hspace{-1cm}\bar{E}_0(1/2) \simeq \frac{1}{2}+ \frac{1}{2}\gamma
+ 4x_Ay^2+4x_Bv^2
+ \hspace{1mm}x_A^2\gamma\Big[x^4 +y^4+ 
8x^2y^2-4x^3y\Big] \nonumber \\ 
&&\hspace{2.3mm}+\hspace{1mm}x_Ax_B\gamma\Big [-8xyuv \left \{
\cos(\theta+\xi)+ \cos(2\theta+\xi)\right \}
-\hspace{1mm}4xyu^2\cos\theta + 
4x^2uv\cos(\theta+\xi) \Big . \nonumber \\
\Big.&&\hspace{2.5cm}+\hspace{1mm}2x^2u^2\cos\theta + 
2y^2v^2\cos(3\theta+2\xi)\Big],
\label{E_0_approx}
\end{eqnarray}
From this we see that the phases $\theta$ and $\xi$ only appear in the 
last term proportional to $x_B$. It is clear that $\bar{E}_0$ 
is stationary with respect to 
these phases when they take the values 0 and $\pi$.
To explore the various possibilities, we define the function
\begin{eqnarray}
\label{4.2.2 f function}
\nonumber
&&\hspace{-2.5cm}f(x,u,\xi,\theta) = 
-\hspace{1mm}8xyuv[\cos(\theta+\xi)+\cos(2\theta+\xi)]-4xyu^2
\cos\theta \\ 
&&\hspace{0.05cm}+\hspace{1mm} 4x^2uv\cos(\theta+\xi) 
+2x^2u^2\cos\theta + 2y^2v^2\cos(3\theta+2\xi),
\end{eqnarray}
which is the quantity multiplying $x_Ax_B\gamma$ in
Eq.~(\ref{E_0_approx}). This function is tabulated in
Table~\ref{table} for various values of $\xi$ and $\theta$. 
\begin{table}
\linespread{1}
\begin{center}
\begin{tabular}{ | l | l | l |}
    \hline
    $\xi$ & $\theta$ & $f(x,u,\xi,\theta)$ \\ \hline
    0 & 0 & $-16xyuv-4xyu^2+4x^2uv+2x^2u^2+2y^2v^2$ \\ \hline
    0 & $\pi$ & $4xyu^2-4x^2uy-2x^2u^2-2y^2v^2$ \\ \hline
    $\pi$ & 0 & $+16xyuv-4xyu^2-4x^2uv+2x^2u^2+2y^2v^2$ \\     \hline
     $\pi$ & $\pi$ & $4xyu^2+4x^2uy-2x^2u^2-2y^2v^2$\\     \hline
    \end{tabular}
    \end{center}
    \caption{The function $f(x,u,\xi,\theta)$ defined in Eq.~(\ref{4.2.2 f function}) tabulated for various 
values of $\xi$ and $\theta$.}
\label{table}
\end{table}
From this table it is clear that $\xi=0$, $\theta=\pi$ will give a 
lower energy than $\xi=\pi$, $\theta=\pi$. For $\xi=0$, $\theta=0$ we 
have
\begin{equation}
\label{4.2.2 f 1}
f(x,u,0,0)-f(x,-u,0,0) = 8xuv(x-4y).
\end{equation}
Since $x_A$ is close to 1, Eq.~(\ref{E_0_approx}) is minimized for a
value of $x$ close to $1/\sqrt{2}$ which is much larger than $y$.
This implies that any minima of the function $f(x,u,0,0)$ will
occur for
\textit{negative} values of $u$. But $u$ must be positive (recall 
$u=|d_0|$), so this case must be rejected.
Finally, for $\xi=\pi$, $\theta=0$, we have
\begin{equation}
\label{4.2.2 f 2}
f(x,u,\pi,0)-f(x,-u,\pi,0) = -8xuv(x-4y).
\end{equation}
The same argument implies that minima of $f(x,u,\pi,0)$ must
occur at
\textit{positive} $u$. We are thus left with the two
possibilities $\xi = 0$, $\theta = \pi$ and $\xi = \pi$ and
$\theta = 0$. A comparison of the contour plots of 
$\bar E_0(x,u,\pi,0)$ and $\bar E_0(x,u,0,\pi)$ shows that 
the latter is the one that 
provides the lowest energy. 
For $x_A=0.95$ and $\gamma=2$, $\bar E_0(x,u,0,\pi)$ is
minimized for $x_{min}\simeq 0.697$ and $u_{min}\simeq 0.677$. 
The value of $x_{min}$ found here is close to the value of 0.696 found 
for $x_A= 1$. Not surprisingly, the $|c_m|^2$ coefficients
are close to the values obtained in the single-species limit.

We will now use the value of $\bar E_0(1/2)$ to show that
persistent currents are possible for $l >1$. To be specific, we
consider $l=1+l^{\prime}$ with $0\leq l^{\prime} \leq 1$. Using the
periodicity of $\bar \epsilon_0(l)$, we have
\begin{equation}
\label{4.2.2. e0 l}
{E}_0(1+l^{\prime}) = 
1+2l^{\prime}+\bar{E}_0(l^{\prime}).
\end{equation}
At $l=x_A$, Eq.~(\ref{4.2. energy simplified}) gives 
$\bar{E}_0(x_A)=x_A+\gamma/2$. We then find that 
$\bar{E}_0(1+x_A)-\gamma/2=1+3x_A=3.85$ for $x_A=0.95$. As
explained earlier, this value is exact within the mean-field 
analysis. We next use Eq.~(\ref{4.2.2. e0 l}) to obtain
\begin{equation}
\bar{E}_0\left(3/2 \right) = 2+ \bar{E}_0\left( 1/2 \right).
\end{equation}
\begin{figure*}[t]
\centering \scalebox{0.4}
{\includegraphics[50,200][575,575]{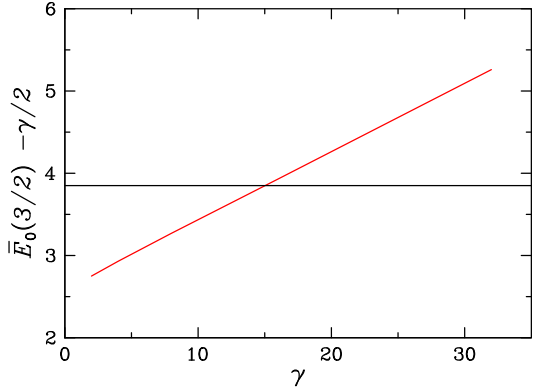}}
\caption{The energy at $l=3/2$ vs $\gamma$ for $x_A = 0.95$. The
horizontal line is the value of $\bar E_0(1+x_A)-\gamma/2$.
}
\label{fig:E_0_vs_gamma} 
\end{figure*}
{\noindent In Fig.~\ref{fig:E_0_vs_gamma} we show }
the behaviour of $\bar{E}_0(3/2)-\gamma/2$ as a function of 
$\gamma$ for $x_A = 0.95$. We see that $\bar E_0(3/2)$ becomes
larger than $\bar E_0(1.95)$ at a value of $\gamma \simeq 15$.
This implies the existence of a local minimum in the range $1.5
< l < 1.95$ and hence the possibility of persistent currents.
The value $\gamma \simeq 15$ is clearly an upper bound to
$\gamma_{cr}$ for this value of $x_A$.

\begin{figure*}[t]
\centering \scalebox{0.4}
{\includegraphics[50,200][575,575]{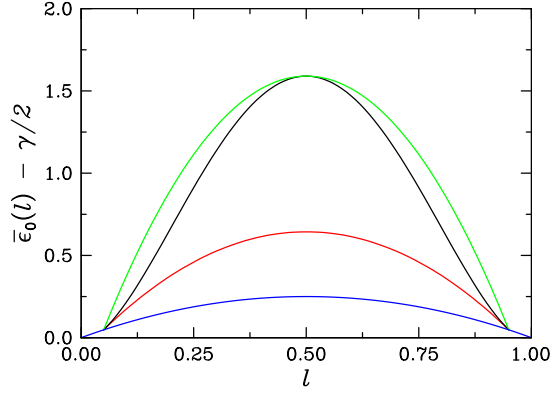}}
\caption{ The function $\bar e_0(l) -\gamma/2$ plotted vs $l$ in
different approximations. The blue curve is the function
$l(1-l)$; the red, black and green curves are $\bar e_0^{(1)}$,
$\bar e_0^{(2)}$ and $\bar e_0^{(3)}$ respectively. $\gamma =16$
and $x_A=0.95$.
}
\label{fig:epsilon} 
\end{figure*}

The approximate behaviour of $\bar E_0(l)$ as a function of $l$
can be obtained by generating approximations to $\bar
e_0(l)$. For $0\le l \le x_B$ and $x_A \le l \le 1$, $\bar
e_0(l) - \gamma /2 = l(1-l)$. From Eq.~(\ref{4.2. E new 5
simple}) we have $\bar
e'_0|_{l=x_A^-} = 1-2x_A - \lambda$. The
simplest approximation to $\bar e_0(l)$ in
the range $x_B\le l\le x_A$ consistent with this information is
\begin{equation}
\bar e_0^{(1)}(l) -\gamma/2 = l(1-l) + \lambda 
{(x_A-l)(l -x_B)\over x_A-x_B}
\end{equation}
An improved approximation is a fit that reproduces the value of
$\bar e_0(l)$ at $l = 1/2$. It takes the form
\begin{equation}
\bar e_0^{(2)}(l) -\gamma/2 = l(1-l) + 
\lambda {(x_A-l)(l -x_B)\over x_A-x_B}
+\mu {(x_A-l)^2(l -x_B)^2\over (x_A-x_B)^4}
\end{equation}
where $\mu =
16[\bar e_0(1/2)-\gamma/2-1/4-\lambda(x_A-x_B)/4]$.
A third approximation ignores the information about the slope of
$\bar \epsilon_0(l)$ at $l=x_A$ but includes the value at
$l=1/2$. This approximation gives
\begin{equation}
\bar e_0^{(3)}(l) -\gamma/2 = l(1-l) + 
\nu {(x_A-l)(l-x_B)\over (x_A-x_B)^2},
\end{equation}
where $\nu = 4(\bar e_0(1/2) -\gamma/2 -1/4)$.
These various approximations are plotted in Fig.~\ref{fig:epsilon}
for $\gamma = 16$. We expect the correct variation of 
$\bar e_0(l)$ to be bounded by the 
$\bar e_0^{(2)}(l)$ and $\bar e_0^{(3)}(l)$ curves; for $l\to
0.95$, $\bar e_0(l)$ should be closer to the $\bar e_0^{(2)}(l)$
curve but for $l\to 0.5$ it should be closer to the 
$\bar e_0^{(3)}(l)$ curve. We note that $\bar e_0^{(3)}(l)$ must 
give the correct behaviour in the $x_A\to 1$ limit. 

\begin{figure*}[!ht]
\centering \scalebox{0.4}
{\includegraphics[50,200][575,575]{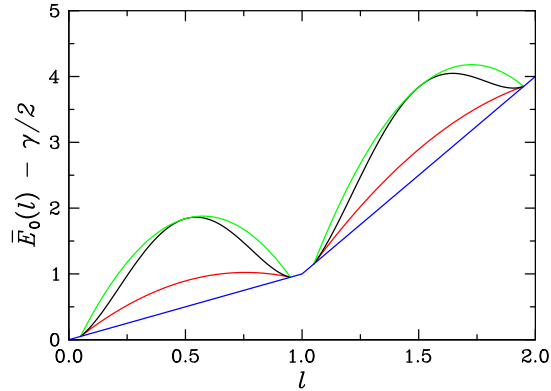}}
\caption{The energy $\bar E_0(l)-\gamma/2$ vs $l$ for $\gamma
=16$ and $x_A=0.95$. The various curves correspond to the
various approximations to $\bar e_0(l)$ shown in
Fig.~\ref{fig:epsilon}.
}
\label{fig:energy} 
\end{figure*}
These different approximations can be used to determine
corresponding approximations to $\bar
E_0(l)$, which is plotted in Fig.~\ref{fig:energy} in the range
$0\le l \le 2$ for $\gamma =16$. The red curve based on
$\bar e_0^{(1)}(l)$ does not show a local minimum at $l=1.95$ as
predicted by considerations of the slope of $\bar E_0(l)$ at this point.
On the other hand, the black curve based on $\bar e_0^{(2)}(l)$
which includes the information about $\bar e_0(1/2)$ shows a
local minimum below $l=1.95$ and demonstrates 
that persistent currents should be possible for $l$ between 1.5
and 2. Regarding persistent currents at $l=0.95$, the critical
interaction strength according to Eq.~(\ref{4.2. gamma cr}) 
for $x_A = 0.95$ is $\gamma_{\rm cr} \simeq 1.9$. For this value
of $\gamma$, $\bar e_0^{(1)}(1/2)$ is actually quite close to 
to the true value $\bar e_0(1/2)$ as determined by the
four-component wave function analysis. Thus the prediction of the
critical interaction strength based on the slope of $\bar
E_0(l)$ remains quite accurate in this case. However, it is
clear that the slope of $\bar{E}_0(l)$ calculated at
$l=(x_A+n-1)^-$, although correct, is not sufficient to provide
a criterion for the existence of persistent currents for $n>1$.
We have also extended the analysis to slightly smaller values of
$x_A$ and arrive at similar conclusions. However, increasingly larger
values of $\gamma$ are then required to achieve a local minimum
between $l=1.5$ and $l=1+x_A$.
 
We finally mention the behaviour of $\bar{E}_0(l=1/2)$ when $x_A=1/2$. 
In this limit the expression for $\bar{E}_0(l)$ given in Eq.~(\ref{4.2. 
energy simplified}) is correct for \textit{all} $l$ and gives in 
particular $\bar{E}_0(l=1/2,x_A=1/2)=1/2+\gamma/2$. This value is 
reproduced by Eq.~(\ref{4.2.2 energy final}) at $x_A=1/2$ 
irrespective of 
the phases $\beta$ and $\xi$ since the minimum occurs for $\theta=\pi$ 
and $x=u=1/\sqrt{2}$, where all the $\beta$ and $\xi$ dependent terms 
have no effect since $y=v=0$. We note that the minimizing value of
$\theta$ is the same as in the $x_A\to 1$ limit and anticipate
that this will remain true for intermediate values of $x_A$ between 
$x_A=1/2$ and $x_A=1$. However, a more careful analysis of 
Eq.~(\ref{4.2.2 energy 
final}) would be required to confirm this and to determine the
remaining variational parameters that minimize the GP energy.

\section{Conclusions}
In this paper we have extended to the two-species system Bloch's 
original argument regarding the possibility of persistent
currents in the idealized
one-dimensional ring geometry. Strict periodicity of the energy
$e_\alpha(L)$ defined in Eq.~(\ref{spec_E}) is found to arise
when the mass ratio $M_A/M_B$ is a rational number. By making a
connection to the Landau criterion for the special case
$M_A=M_B$, we show that persistent currents are in general
possible at the discrete set of total angular momenta $L_n =
nN\hbar$, except when
the interaction parameters satisfy the condition in
Eq.~(\ref{U_ij_equality}). The underlying reason for this
limitation is the existence of excitations with a 
particle-like dispersion. This
conclusion is consistent with the predictions of a mean-field
analysis based on the GP energy functional. A detailed analysis
of the GP energy in the vicinity of $l = x_A$, first carried out by
Smyrnakis {\it et al.}~\cite{Smyrnakis09}, indicates that persistent
currents are possible at this angular momentum per particle if
the interaction parameter exceeds the critical value given in
Eq.~(\ref{4.2. gamma cr}). These authors go on to
claim that persistent currents cannot arise for $l>1$ in
the two-species system. However, a more detailed analysis of the
global behaviour of the GP energy demonstrates that
this conclusion is not valid. Quite generally, the properties of
the two-species system evolve continuously to those of the
single-species system as the concentration of the minority
component is reduced. It would of course be of interest to
verify these theoretical predictions experimentally. The recent
experimental realization of toroidal Bose-Einstein
condensates~\cite{Ryu07,Ramanathan11} would suggest that
experiments on two-species systems may soon be feasible.

\acknowledgments
This work was supported by a grant from the Natural Sciences and
Engineering Research Council of Canada.

\end{document}